\documentclass[10pt,conference]{IEEEtran}
\IEEEoverridecommandlockouts

\usepackage{booktabs}
\usepackage{cite}
\usepackage{framed}
\usepackage[hidelinks]{hyperref}
\usepackage{xspace}
\usepackage{svg}
\usepackage{subcaption}
\usepackage{tabularx}
\usepackage[]{todonotes}
\usepackage{comment}
\usepackage{url}
\usepackage{xcolor}
\def\BibTeX{{\rm B\kern-.05em{\sc i\kern-.025em b}\kern-.08em
    T\kern-.1667em\lower.7ex\hbox{E}\kern-.125emX}}

\newcolumntype{Y}{>{\raggedright\arraybackslash}X}
\newcolumntype{A}[1]{>{\raggedleft\arraybackslash}p{#1}}

\newcommand{\etal}{\textit{et al.}}
\newcommand{\mcp}{MCP\xspace}
\newcommand{\llm}{LLM\xspace}
\newcommand{\github}{GitHub\xspace}
\newcommand{\eg}{e.g.,\xspace}
\newcommand{\account}[1]{\emph{#1}}
\newcommand{\fig}[1]{Fig.~\ref{#1}}
\newcommand{\tab}[1]{Table~\ref{#1}}
\newcommand{\sect}[1]{Section~\ref{#1}}

\begin{document}


\title{A Two-Dimensional Study of the Model Context Protocol: Publication and Adoption}



\author{
\IEEEauthorblockN{Natarajan Chidambaram}
    \IEEEauthorblockA{
        \textit{Luxembourg Institute of Science and Technology}\\
        Esch-sur-Alzette, Luxembourg \\
        natarajan.chidambaram@list.lu}
\and
\IEEEauthorblockN{Mauro Dalle Lucca Tosi}
    \IEEEauthorblockA{
        \textit{Luxembourg Institute of Science and Technology}\\
        Esch-sur-Alzette, Luxembourg \\
        mauro.dalle-lucca-tosi@list.lu}
\and
\IEEEauthorblockN{\centerline{Jordi Cabot}}
    \IEEEauthorblockA{
        \textit{\centerline{Luxembourg Institute of Science and Technology}}\\
        \centerline{and \textit{University of Luxembourg}}\\
        \centerline{Esch-sur-Alzette, Luxembourg} \\
        \centerline{jordi.cabot@list.lu}}
}

\maketitle

\begin{abstract}
The Model Context Protocol (\mcp), released by Anthropic in November 2024, standardizes how large language model applications connect to external tools and data sources. 
Despite MCP's rapid growth, no study has jointly characterized its emergence in the research literature, its adoption on \github and the relationship between them. 
We address this gap with a longitudinal, two-dimensional study of 802 \mcp-related publications and 33,319 \github repositories.
We characterize their growth and identify their application domains via BERTopic and usage contexts via a human-validated GPT-5.6-Luna classifier.
We find that publications and repositories began rapid growth within a month of each other in 2025, and peaked together in March 2026. 
Notably, a second growth wave alone accounts for 49.5\% of the GitHub repositories. 
Most publications (57.8\%) and nearly all repositories (93.7\%) use MCP as an enabling technology rather than for analyzing, evaluating, extending, or securing it directly.
Among repositories, this latter category concentrates almost entirely in one general-purpose topic, while applied usage spreads into product-specific niches. 
Finally, we provide an overview of the MCP landscape and recommendations to researchers and practitioners in this field. 
\end{abstract}


\section{Introduction}
\label{sec:introduction}

The Model Context Protocol (\mcp) was released publicly on 25 November 2024 as an open protocol to standardize how large language model (\llm) applications connect to external tools and data sources~\cite{antropic2024}.
Within the first year, Anthropic reported over 97M monthly SDK downloads and 10K active servers, with first-class client support from major AI platforms such as ChatGPT and Cursor~\cite{parra2025mcp}.
Understanding this fast-growing ecosystem can provide a baseline for protocol maintainers, researchers investigating narrower questions such as security or maintainability, and tool builders and educators deciding which application domains to target. 

Several recent studies have examined specific aspects of \mcp's \github ecosystem, such as server security, maintainability, and construction patterns~\cite{hasan2026security, lin2025evolvable, mastouri2026api, toeppe2026dataset}, but none track adoption over time or across application domains.
However, to our knowledge, \mcp's presence in academic publications remains uncharacterized.

In other domains, previous analyzes have examined academic publications and \github repositories in parallel to identify the differences between research and usage of \github automation tools~\cite{saroar2024marketplace} and to study academic and developer perspectives in fixing vulnerabilities~\cite{salzano2025bridging}. 
Building on this perspective, we investigate \mcp through its emergence in the research literature and its presence in the \github ecosystem, tracking how research attention and MCP-related repositories evolve and examining their relationship.

We address this gap with a longitudinal, two-dimensional empirical study.
We construct and independently analyze a dataset comprising 802 \mcp-related publications from Lens.org and 33,319 \github repositories, capturing repositories beyond established \mcp implementations.
We characterize \mcp's emergence in terms of its growth over time, application domains and usage contexts.
We then compare both dimensions to determine whether research attention and \github activity exhibit similar or distinct patterns in timing, application domains and usage contexts. 
In this paper, we specifically answer the following research questions:

\indent \textbf{RQ1: How has \mcp emerged in research and practice?} 
We examine the temporal growth in the number of \mcp-related academic publications (RQ1.1) and \github repositories (RQ1.2), and how these trajectories relate (RQ1.3).

\indent \textbf{RQ2: What are the predominant application domains and usage contexts of \mcp in research and the \github ecosystem?}
We examine this separately for \mcp-related academic publications (RQ2.1) and \github repositories (RQ2.2), and compare the two (RQ2.3).

Our analysis reveals two main findings. First, despite the substantial difference in scale between \mcp-related academic publications and \github repositories, their growth is closely synchronized. 
They peak in the same month in 2026, with both dimensions still growing.
Second, most academic publications (57.8\%) and the large majority of \github repositories (93.7\%) use \mcp as an enabling technology rather than studying it directly, though this skew is far more pronounced on \github. 
Repositories that study, evaluate, extend, or secure \mcp itself concentrate almost exclusively in a single general-purpose \mcp tooling topic, while repositories that apply \mcp diversify into product-specific niches such as email access, trading, and biomedical tooling.

\sect{sec:related_work} reviews related work, and \sect{sec:dataset} explains how we constructed the dataset. In \sect{sec:findings}, we report our findings for RQ1 and RQ2 separately before examining how their results relate to one another. \sect{sec:discussion} discusses cross-cutting implications, and \sect{sec:threatstovalidity} reports threats to validity. Finally, \sect{sec:conclusion_future} concludes the paper. A replication package is present in \url{https://doi.org/10.5281/zenodo.22705762}.

\section{Related work}
\label{sec:related_work}

To comprehensively analyze the evolution and adoption of MCP, we consider two complementary facets: its development within the scientific community, as captured through scientometric and metascience analyzes, and its uptake in open-source software, as observed through mining \github repositories. 
We also examine how these two facets evolve in relation to one another.

\subsection{On metascience studies on MCP}

Metascience and scientometric analyses are widely used to study the development of research fields. By examining publication growth, research topics, application domains, and other metadata, these studies provide a structured view of how scientific attention evolves~\cite{demetrescu2022computer,dalle2022understanding, de2023global}. Comparable systematic mapping studies are also common in software engineering, where they are used to classify the literature, identify research trends, and reveal gaps between topics~\cite{tosi2024metascience,tosi2025ocl}.

To the best of our knowledge, no such study has systematically analyzed the peer-reviewed literature on MCP. 
While MCP-related research is growing, existing work predominantly focuses on proposing, evaluating, or applying the protocol rather than characterizing the evolution and domain distribution of the research field itself. This paper fills this gap by systematically characterizing the evolution of MCP publications, the venues in which they are published, and their distribution across application domains and usage contexts.

\subsection{On mining GitHub repositories}

Empirical studies on \mcp's \github ecosystem have been published within about a year of the protocol's release, spanning dataset construction and characterization of \mcp servers~\cite{toeppe2026dataset, hasan2026security, lin2025evolvable, mastouri2026api}.

Toeppe et al.~\cite{toeppe2026dataset} present the first large-scale, evidence-based dataset of \mcp implementations collected directly from \github, constructing 2,297 verified repositories.
By analyzing dependency files, executable entry points, and deployment-environment signals, they verified and classified each repository as a working server, client, or gateway, finding a median of four active contributors per repository.
Lin et al.~\cite{lin2025evolvable} relied on mcp.so
to build a structured dataset of 14K \mcp servers and clients.
For the artifacts with an associated \github repository, they complemented the data obtained from mcp.so with repository-level metadata such as stars and forks, 
and categorized each artifact by its type (server or client) and application category (cloud, AI, data).
They found a majority of artifacts have fewer than ten stars.
Hasan et al.~\cite{hasan2026security} present the first large-scale empirical study of \mcp health, security, and maintainability of 1,899 open-source \mcp servers, applying a 10-star popularity filter, they found 583 non-toy repositories.
Relative to traditional open-source software, these \mcp servers show higher development activity (\eg commits/week) and greater community reach (\eg stars gained/year), while exhibiting comparable maintainability issues (\eg code smells).
In addition, they highlighted the need for \mcp-specific vulnerability detection techniques, as not all were captured with existing patterns.
%
Mastouri et al.~\cite{mastouri2026api} present the first large-scale empirical study of \mcp server construction.
By analyzing 116 official \mcp servers (with at least 10 stars), they found 89\% are fully or partially backed by an existing REST API, with 92\% of tools implemented as bare API wrappers exposing a median of only 19\% of available operations.
 
We can clearly observe that these datasets are scoped to established \mcp\ implementations, verified directly~\cite{toeppe2026dataset} or listed on registries~\cite{lin2025evolvable}.
Similar empirical analyses on \mcp servers, such as developer-reported challenges in \mcp adoption~\cite{oyelayo2026challenges}, taxonomies of \mcp server faults~\cite{taraghi2026faults} and analyzing AI agent tools created from \mcp server repositories~\cite{stein2026tools} continue to grow.
However, none of these studies, nor the datasets described above, track how repository-level adoption evolves, categorize adopted repositories by application domain and usage context in a way that could be related to the concurrent research literature on \mcp, or include repositories that engage with \mcp without having \mcp at their core focus.
This paper addresses both gaps directly.

\subsection{On paper-repository linkage}

Previous work has explicitly connected academic publications and \github repositories in either direction.
Escamilla \etal~\cite{escamilla2022rise} identified Git hosting platform URIs in a large scholarly dataset, whereas Wattanakriengkrai \etal~\cite{wattanakriengkrai2022github} identified references to academic papers in \github repositories.
Alrashedy and Binjahlan~\cite{alrashedy2024software} extracted and classified \github links from software-engineering publications to study research artifacts, while Fan \etal~\cite{fan2021makes} linked AI papers to accompanying repositories to characterize features associated with repository popularity.

Related work has also independently characterized academic research and open-source practice within a common topic.
Saroar \etal~\cite{saroar2024marketplace} compared software-automation literature with \github Marketplace tools, identifying differences between research attention and developer adoption.
Similarly, Salzano \etal~\cite{salzano2025bridging} compared vulnerability-fixing strategies derived from the literature with those observed in \github repositories, revealing both agreement and practices absent from literature.

Overall, existing work links individual papers and repositories, or compares research and practice at a given point in time.
However, to our knowledge, no work has examined whether the joint evolution of publications and repositories in a specific emerging area can provide ecosystem-level insights.
This paper, rather than tracing scientific publications to their repositories, tracks the volume and application-domain distributions of \mcp-related publications and \github repositories, and compares their evolution to identify convergences and gaps between scientific attention and practical adoption.
\section{Dataset}
\label{sec:dataset}

\subsection{Research publications}

To construct a comprehensive dataset of publications related to MCP, we applied the following six-step procedure:

\begin{enumerate}
    \item \textbf{Data collection}: On June 8, 2026, we queried Lens.org~\cite{lens2026scholarly}, a platform aggregating scholarly metadata from sources such as ORCID, CrossRef, and PubMed. We retrieved publications containing \texttt{Model Context Protocol} or its lemmatized form in the title, abstract, field of study, or keywords. This search returned 1,185 MCP-related publications.

    \item \textbf{Duplicate removal}: We manually inspected the records and identified 224 publications with one or more duplicates. These duplicate records were identified and removed based on identical titles and author lists, alongside one article with a broken DOI that could not be located online. When selecting among duplicate records, preference was given to formally published scholarly versions, including journal and conference papers, over code repositories or dataset records. Where multiple versions of the same work were available, the earliest version was retained to reflect the point in time at which the research was first released. This approach resulted in a final set of 879 unique publications.

    \item \textbf{Publication-type correction}: As publication types were occasionally incorrectly indexed (e.g., conference proceedings classified as book chapters or journal articles), we manually reviewed and corrected the publication type of 95 records. These corrections included the reclassification of book chapters as conference proceedings, journal articles as preprints or conference proceedings, and reports as preprints.

    \item \textbf{Publication-type selection}: To reduce noise in the subsequent analysis, we excluded 66 records that were classified as \textit{other} (\eg blog posts, slides, code), 3 dissertations, and 8 datasets. Datasets and code repositories were excluded because in this analysis we focus on text-based publications. After this filtering step, the dataset comprised 802 publications.

    \item \textbf{Venue-name normalization}: We standardized venue names by removing edition-specific information from journals and using acronyms for conferences, allowing the same venue to be identified across editions.

    \item \textbf{Filling missing information}: We searched for missing publication dates for 100 records and missing abstracts for 69 records. This included articles without a publication month, as well as those dated January 1, which is commonly used as a default when the specific month of publication is unavailable.

\end{enumerate}

Overall, this procedure resulted in a dataset of 802 unique MCP-related publications. For each publication, the publication type, venue name, and publication date were manually reviewed and, where necessary, corrected.

\subsection{GitHub repositories}
\label{sec:github_repo_data}
\paragraph{\textbf{Repository identification}}
To obtain MCP-related GitHub repositories, we relied on the \github REST API's Search endpoint~\footnote{\url{https://api.github.com/search/repositories}}. 
Querying this endpoint on 14 June 2026 with 24 lexical and orthographic variants of MCP terminology such as \texttt{model context protocol}, \texttt{mcp server}, \texttt{mcp client}, \texttt{mcp plugin}, \texttt{fastmcp}, etc., yielded a set of non-archived repositories as shown in \tab{tab:funnel}.
We deliberately excluded \texttt{mcp} as it would fetch repositories related to \textit{Mod coder pack}. 

As the search endpoint caps any single query at 1,000 repositories, we applied recursive partitioning on the number of stars and the repository creation date until every query fell under the cap. 
For example, \texttt{mcp-server} matched 76,249 repositories, where the repositories with zero stars alone required 14 subdivisions on repository creation date before the earliest query returned fewer than 1,000 repositories. 
For each repository, we extracted its metadata such as contributor count, commit count, raw README content, description, repository creation date, last pushed date and so on.
\begin{table}[t]
\caption{Number of repositories retained and removed at each filtering step.}
\label{tab:funnel}
\centering
\small
\begin{tabular}{l|rr}
\textbf{Criterion} & \textbf{Retained} & \textbf{Removed} \\
\midrule
initial \#non-archived repositories & 325,580 & --- \\
unique repositories  & 148,919 & --- \\
non-empty repositories  & 144,626 &  4,293 \\
at least one commit after creation            & 143,694 &    932 \\
has a code file          & 132,153 & 11,541 \\
active on or after 1 January 2025        & 131,727 &    426 \\
$\geq$3 stars$|$$\geq$3 contributors$|$$\geq$50 commits
                                                 & \textbf{33,319} & 98,408 \\
\end{tabular}
\end{table}

\begin{figure}
    \centering
    \includegraphics[width=\columnwidth]{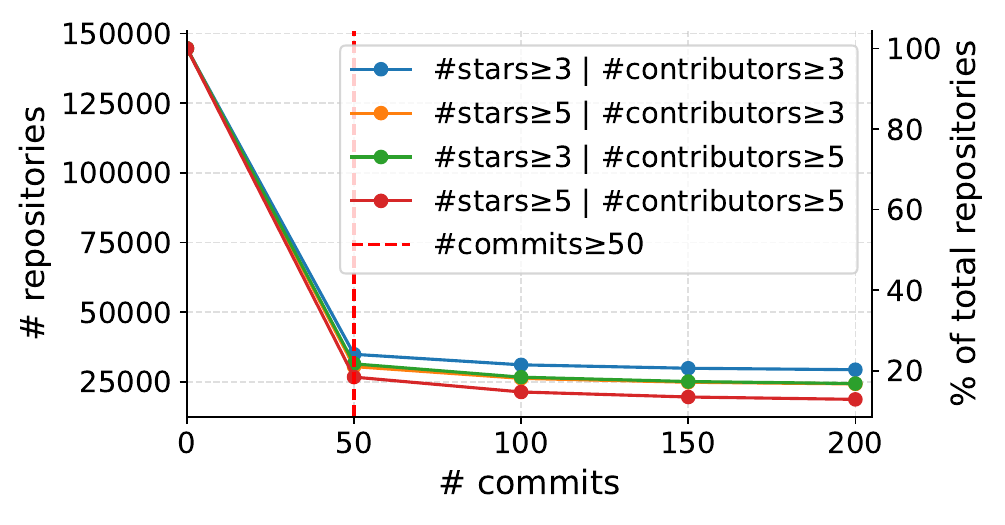}
    \caption{Sensitivity of dataset size to number of stars, contributors and commits. The red dashed line is the selected commits threshold.}
    \label{fig:threshold}
\end{figure}

\paragraph{\textbf{Repository filtering}}
To remove toy and experimental repositories and identify candidate repositories for our analysis, we relied on the heuristic that a repository should not be empty (\eg \textit{MacStenk/semantic-search-mcp}) and should satisfy activity and metadata criteria, such as being active on or after 1 January 2025 and having at least one code file~\cite{munaiah2017curating, toeppe2026dataset}.
\tab{tab:funnel} reports the number of repositories retained and removed at each filtering step. 
As a final criterion, we considered a repository meeting any of three distinct signals: at least three stars, three contributors, or 50 commits. 
We used disjunctions rather than conjunctions, as these three capture different signals, namely community interest~\cite{borges2018star}, collaboration, and development volume~\cite{munaiah2017curating}. 
Requiring all three simultaneously would retain only 3,703 repositories and would remove single-maintainer projects with substantial stars and commits such as \textit{alefcarlos/mcpserver-auth-demo}.
On one hand, we set the contributor threshold at three rather than two, since two contributors could already be satisfied by a single human account alongside an automated account (e.g., GitHub Actions bot) such as \textit{Mugetsu44-44/mcp-gateway-orchestrator} 
rather than genuine collaboration.
On the other hand, adopting a stricter alternative would discard more repositories without a proportional gain in excluding toy repositories.
\fig{fig:threshold} reports the sensitivity of dataset size to the commit-count threshold, across four star/contributor threshold combinations, and the final commit threshold of 50 is motivated by the observed knee bend~\cite{satopaa2011kneedle}.

To assess the precision of the heuristics, we randomly selected and manually inspected a sample of 380 repositories from the final set of 33,319 repositories. 
We determined the sample size using Cochran’s formula with finite population correction, assuming a 95\% confidence level and a 5\% margin of error~\cite{cochran1977sampling}. 
We then classified each repository as genuine or toy based on their contents independent of the heuristics under evaluation.
Two authors of this paper labeled the sample independently, reaching an agreement on 97.8\% of repositories with Cohen's $\kappa$=0.66~\cite{jacob1960agreement} (substantial agreement) and Gwet's AC$_{1}$=0.98~\cite{gwet2008agreement} (almost perfect agreement). 
We report AC$_{1}$ alongside $\kappa$ because the sample is skewed towards genuine repositories, a condition under which $\kappa$ is known to understate agreements~\cite{feinstein1990high}.
Disagreements were resolved through discussion, yielding 96.8\% genuine repositories in the sample (no repositories were removed from the dataset on this basis).

\section{Findings}
\label{sec:findings}

\subsection{\textbf{RQ1}: How has MCP emerged in research and practice?}
\label{sec:rq1}

This research question investigates \mcp's growth over time, in terms of the number of research publications and \github repositories since its release.

\subsubsection{\textbf{RQ1.1}: How has academic interest in \mcp evolved over time?}
\label{sec:rq11}

\begin{figure}
    \centering
    \includesvg[width=\linewidth]{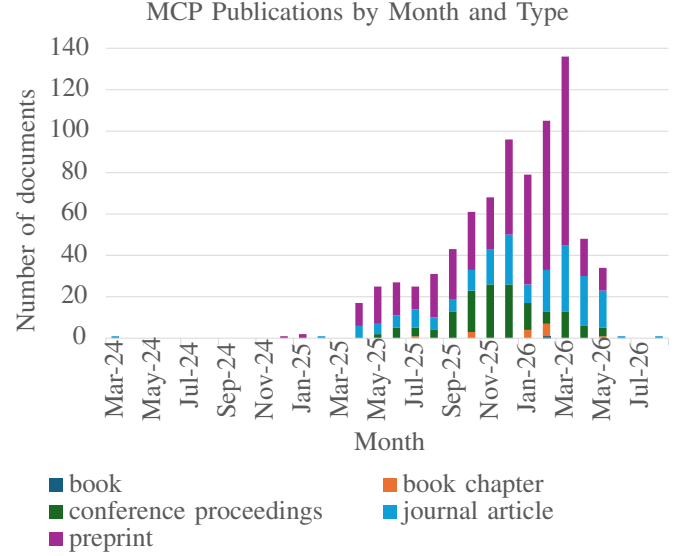}
    \caption{Monthly number of MCP-related publications, disaggregated by publication type.}
    \label{fig:publication_number}
\end{figure}

To examine the evolution of academic interest in MCP, we analyzed the monthly number of MCP-related publications by publication type. As shown in \fig{fig:publication_number}, publication activity remained limited in 2024 and began to increase noticeably from April 2025. This increase is primarily driven by preprints, although the number of journal articles also grows over the period considered.

Conference-proceedings publications started to increase from September 2025 with 13 records and held 30\% of publications for that month, before declining to 9\% from February to May 2026. Such decrease is likely explained by indexing lag, since conference papers may be published several months after the event. As the dataset was collected on 8 June 2026, some papers presented between April and June 2026 may not yet have been published and therefore may not be included in our analysis. The earliest MCP-related publication identified in the dataset was published on 30 March 2024~\cite{Gaddam_2024}.

\begin{figure}
    \centering
    \includesvg[width=\linewidth]{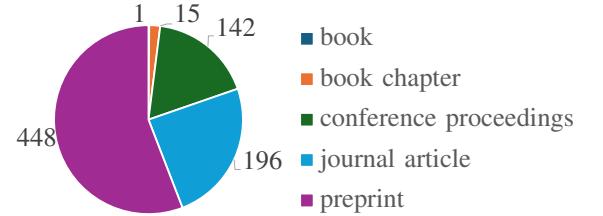}
    \caption{Overall distribution of MCP-related publications by publication type.}
    \label{fig:publication_type}
\end{figure}

\fig{fig:publication_type} presents the overall distribution of \mcp-related publications by publication type. 
Preprints account for the largest share of the dataset, with 448 publications (56\%). This predominance is unsurprising, given the novelty of \mcp research and the role of preprints in rapid dissemination of results. Journal articles constitute 196 publications (24\%), while conference proceedings account for 142 publications (18\%). The remaining publications comprise 15 book chapters (2\%) and one book.

To investigate where \mcp-related research is published, we further analyzed the number of publications per conference and journal. \tab{tab:mcp-venues} lists the venues that have published \mcp-related articles. The two venues with the most publications are AAAI (7 conference publications) and IJSREM (8 journal publications).

\begin{table}[t]
    \centering
    \small
    \caption{Distribution of MCP-related publications by venue. 
    Venues with the same number of publications are grouped.}
    \label{tab:mcp-venues}
    \textbf{(a) Conferences}\par\smallskip
    \begin{tabularx}{\columnwidth}{@{}Y r@{}}
        \toprule
        \textbf{Conference(s)} & \textbf{\#Publications} \\
        \midrule
        AAAI             & 7 \\
        EMNLP            & 4 \\
        AIAA             & 3 \\
        11 conferences   & 2 \\
        106 conferences  & 1 \\
        \bottomrule
    \end{tabularx}

    \vspace{0.8em}

    \textbf{(b) Journals}\par\smallskip
    \begin{tabularx}{\columnwidth}{@{}Y r@{}}
        \toprule
        \textbf{Journal(s)} & \textbf{\#Publications} \\
        \midrule
        IJSREM & 8 \\
        JISEM & 7 \\
        IJRASET & 6 \\
        Buildings & 5 \\
        IJCTT, IJISRT, WJAETS & 4 \\
        Appl.\ Sci., Information, IJAIBDCMS, IJIRMPS, ISJEM, JCSTS & 3 \\
        18 journals & 2 \\
        104 journals & 1 \\
        \bottomrule
    \end{tabularx}
\end{table}

Most venues in \tab{tab:mcp-venues} are associated with computer science, reflecting the technical origins of \mcp, though the results also suggest interdisciplinary dissemination. 
For instance, the AIAA SciTech Forum, which focuses on aerospace research, contains three \mcp-related publications, while the \textit{Buildings} journal contains five. In addition, the \textit{International Journal of Innovative Research in Engineering \& Multidisciplinary Physical Sciences} (IJIRMPS) and the \textit{International Scientific Journal of Engineering and Management} (ISJEM) have each published three \mcp-related articles. These results indicate that \mcp is attracting interest beyond the computer-science context, such as aerospace, engineering, and applied-science domains.

\begin{framed}
\noindent \textbf{Finding 1.1:} 
Academic interest in MCP has grown sharply since early 2025, driven mainly by preprints, with journal articles and conference proceedings making up a smaller but rising share. Publication venues skew toward computer science, but genuine reach into other domains such as aerospace and engineering.
\end{framed}

\subsubsection{\textbf{RQ1.2}: To what extent has \mcp gained traction among GitHub repositories over time?}
\label{sec:rq12}


\begin{figure}
    \centering
    \includegraphics[width=0.9\columnwidth]{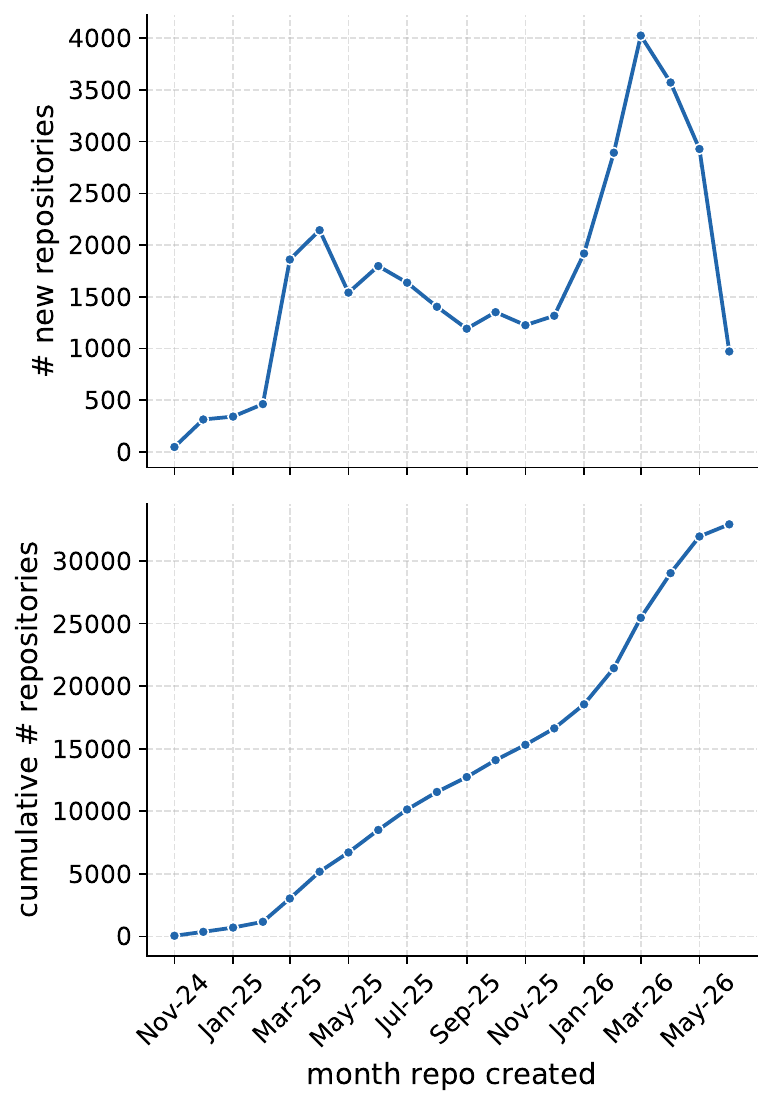}
    \caption{New (top) and cumulative (bottom) number of MCP-related repositories created per month.}
    \label{fig:cumulative_new_repositories}
\end{figure}
We did a monthly aggregation of the repositories in the final dataset (\tab{tab:funnel}) using their creation date (\texttt{created\_at}) and reported the results in \fig{fig:cumulative_new_repositories}. 
The upper panel shows new mature repositories per month and the lower panel shows their cumulative sum.
For the 32,935 repositories (98.85\%) that were created after \mcp's announcement, we use their creation date as a proxy for when their engagement with \mcp likely began, as it is the earliest per-repository timestamp available in our data. 
However, repository creation date cannot serve as this proxy for the 384 repositories (1.15\%) created before \mcp's launch. 
These repositories necessarily existed for a different purpose before the protocol's release, and their true adoption date is not recoverable from creation date alone.
We therefore report these latter repositories separately as a one-off head-start count.

From \fig{fig:cumulative_new_repositories}, we can observe that the growth occurs in two waves rather than a single acceleration. 
A first wave, accounting for 13.6\% of the entire dataset, starts from February 2025 (1,163), is followed by a sharp jump in March 2025 (1,860), and a local peak in April 2025 (2,144) before stabilizing for the remainder of 2025, fluctuating between 1,190 and 1,800 new repositories monthly.

A second wave, accounting for 49.5\% of the entire dataset, starts with sharper acceleration from January 2026 (1,918) to the peak in March 2026 (4,026), visible as the steepest segment of the cumulative curve. 
Then the monthly counts decline at 15\% per month, faster than the first decline at 11\% per month.
This is likely due to the fact that the newer repositories have not yet had enough time to clear the heuristics (\sect{sec:github_repo_data}). This second wave follows two \mcp project milestones in quick succession:
\mcp's donation to the Agentic AI Foundation (9 Dec 2025)~\cite{parra2025mcp} and the launch of MCP Apps (26 Jan 2026)~\cite{mcp_apps_2026}.
These observations are reported descriptively, and we do not infer causation.

\begin{framed}
\noindent \textbf{Finding 1.2:} \mcp-related repository creation grew in two distinct surges.
First, from February 2025 to April 2025, with 13.6\%, and the second from January 2026 to June 2026 accounting for 49.5\% of the dataset.
\end{framed}

\subsubsection{\textbf{RQ1.3}: What relationship exists between the temporal trends observed in academic publications and \github repositories?}
\label{sec:rq13}

The two dimensions in \sect{sec:rq11} and \sect{sec:rq12} show a closely aligned trajectory.
Repository creation jumps in March 2025, and publication output begins increasing in April 2025, placing both inflection points within about a month of each other. 
Growth in both dimensions then peaks in March 2026 with 4,026 new repositories (upper panel of \fig{fig:cumulative_new_repositories}) and 136 publications (\fig{fig:publication_number}), the highest counts observed in either series.
The decline in both series from April 2026 are likely due to the data collection date as discussed in \sect{sec:rq11} and \sect{sec:rq12}.

\begin{framed}
    \noindent\textbf{Finding1.3:} Repository creation and publication output exhibit inflection points approximately a month apart in 2025 and both peak in March 2026. This indicates closely synchronized growth, despite their differing scale.
\end{framed}
\subsection{\textbf{RQ2}: What are the predominant application domains and usage contexts of \mcp in research and practice?}
\label{sec:rq2}
This research question examines the application domains in which \mcp is investigated or applied, as well as the roles that \mcp plays in academic publications and GitHub repositories. 

\subsubsection{\textbf{Experimental setup}}
\label{sec:experimental_setup}
Below we describe the \mcp application domains and usage contexts. \\
\noindent{\textbf{Application domains}}: Since we did not find a predefined taxonomy of \mcp application domains in our literature, we applied the \emph{BERTopic} model~\cite{grootendorst2022bertopic} (a transformer-based semantic topic modeling) separately to the abstracts of the academic publications and the README and description content of each repository \github, fitting an independent model to each. 
The same technique is used in empirical software engineering studies on research-paper abstracts~\cite{wang2024identifying,liu2024unveiling} and other GitHub-derived text~\cite{zhao2024empirical,alam2026analyzing}.
For repository descriptions and README files specifically, our choice of textual input follows prior work that mines a repository's documentation for topic classification~\cite{izadi2021topic}.

We first excluded non-English records, identified using \emph{langdetect}~\cite{nakatani2010langdetect},
and preprocessed the remaining texts by removing formatting artifacts, such as markup and encoding errors.
Next, we generated semantic embeddings with \emph{all-mpnet-base-v2} model~\cite{reimers2019BERT}, a sentence-transformers model fine-tuned from Microsoft's pretrained \textit{microsoft/mpnet-base} checkpoint~\cite{song2020mpnet} and recommended by BERTopic. 
We reduced the embeddings to five dimensions using Uniform Manifold Approximation and Projection (UMAP), configured with cosine distance and 15 neighbors~\cite{mcinnes2018umap}.
Then used \emph{BERTopic}, which clusters document embeddings with HDBSCAN~\cite{campello2013hdbscan} and labels each cluster with class-based TF-IDF (c-TF-IDF) over the cluster's documents. 
We set the minimum topic size to 10 documents for the publication dataset and 200 for the substantially larger repository dataset to limit excessive topic fragmentation. 
Finally, we manually assigned a high-level descriptive label to each cluster based on its central terms.

\noindent{\textbf{MCP usage contexts:}} To identify MCP usage contexts, we classified each publication and repository by the role MCP plays. 
Given the dataset size, we automated this classification using an LLM as a zero-shot classifier, an approach previously shown effective for text classification~\cite{gilardi2023chatgpt}.
We developed the classification taxonomy iteratively. 
The authors manually analyzed a subset of publication abstracts and repository descriptions/READMEs, defined an initial set of categories, and refined them for clarity and mutual exclusivity.
The final taxonomy is presented in Table~\ref{tab:mcp-usage-taxonomy}.

\begin{table*}[t]
\centering
\small
\caption{Taxonomy used to classify the usage context of \mcp in academic publications and GitHub repositories.}
\label{tab:mcp-usage-taxonomy}
\begin{tabularx}{\textwidth}{@{}>{\ttfamily}p{0.28\textwidth} >{\raggedright\arraybackslash}X@{}}
\toprule
\textbf{Category} & \textbf{Description} \\
\midrule

proposes\_extends &
Introduces \mcp itself or proposes a new \mcp-related protocol, specification, or extension. \\
\hline
studies\_analyzes &
Empirically studies, measures, characterizes, explains, describes, or surveys \mcp, including its landscape, adoption, or behavior, without primarily proposing, extending, or securing it. This category also includes educational material about \mcp. \\
\hline
evaluates\_benchmarks &
Develops or applies a benchmark, dataset, testbed, or comparative evaluation for \mcp or \mcp-based systems. \\
\hline
security\_privacy &
Focuses on the security or privacy of \mcp itself, including attacks against \mcp clients or servers, tool poisoning, protocol vulnerabilities, threat models, and related defences. This category does not include projects that merely use \mcp to build a security or privacy application. \\
\hline
uses\_applies &
Uses \mcp as an enabling technology to develop an agent, tool, workflow, or application, or to address a domain-specific problem. \\
\hline
mentions\_only &
Mentions \mcp only peripherally, without making it a substantive focus of the publication or repository. \\

\bottomrule
\end{tabularx}
\end{table*}

Each prompt included the taxonomy and the record's relevant text, namely the title and abstract for publications, and name, description, and README for repositories, and instructed the LLM to assign the best-fitting category.
To assess the reliability of the classification procedure, two authors of this paper independently annotated statistically representative samples from both datasets with 95\% confidence and a 5\% margin of error. 
In total, we manually annotated 259 academic publications and 380 GitHub repositories. 
Cohen's $\kappa$ between each annotator was 0.63 for publications and 0.53 for \github repositories. These values indicate \textit{substantial} agreement for publications and \textit{moderate} agreement for repositories. The authors subsequently discussed and resolved disagreements to construct a ground-truth dataset, which was then used to evaluate the LLM-based classification.


When evaluated against the ground-truth dataset, GPT-5.6-Luna achieved an accuracy of 83.3\% for publications and 93.7\% for GitHub repositories. 
The remaining classification differences should be interpreted in light of the inherently interpretative nature of the taxonomy, particularly where records may plausibly fit more than one category or provide limited contextual information. Cohen's $\kappa$ between GPT and annotators (first, second and ground truth) was 0.63, 0.68, and 0.73 for publications, and 0.51, 0.53 and 0.55 for repositories. All values indicate \textit{substantial} agreement for research papers and \textit{moderate} agreement for \github repositories. The larger gap between raw accuracy and $\kappa$ for repositories reflects that the data is skewed towards a particular category. Notably, the level of agreement between GPT and the annotators was comparable to the agreement observed between the two annotators themselves.
Overall, these results support the use of GPT-based classification for the large-scale analysis presented in the remainder of this study.

\subsubsection{\textbf{RQ2.1}: How are application domains and usage contexts distributed across MCP-related academic publications?}
\label{sec:rq21}

To identify the application domains addressed in MCP-related academic publications, we excluded 29 publications whose abstracts were not written in English, resulting in a final dataset comprising 773 English-language abstracts. Then, we applied the topic modeling previously explained.

As shown in Table~\ref{tab:paper_words}, BERTopic identified six topics and an outlier cluster that groups documents assigned to topics containing fewer than 10 publications. Among the 10 central terms provided by BERTopic, we report five that best represents the cluster. The largest topic is characterized by broad terms related to AI agents and tools,
capturing the general technical literature surrounding MCP and agentic systems.

The remaining topics represent more specific application domains. Two topics, each comprising 42 publications, concern biomedical research and engineering-related domains, respectively. The latter includes research related to simulation, energy, buildings, and construction. This result is consistent with the presence of MCP-related publications in venues such as \textit{Buildings} and the AIAA SciTech Forum. A further topic focuses on healthcare and clinical applications. Another topic is associated with laboratory and scientific-research settings, including applications involving laboratory systems and scientific instrumentation. Finally, a topic of 21 publications concerns financial and blockchain-related applications. The outlier cluster contains publications that could not be reliably assigned to one of the identified topics.


\begin{table}[t]
    \caption{Topics of MCP academic publications, 5 manually selected central terms, and the number of publications in each topic.}
    \label{tab:paper_words}
    \centering
    \footnotesize
    \begin{tabular}{p{2.5cm}|p{4.4cm}|p{1cm}}
        \textbf{Topic} & \textbf{Central terms} & \textbf{\#Publi.}  \\
        \hline
        Agentic systems & tool, ai, agent, systems, language & 588\\
        \hline
        Biomedical & computational, biomedical, drug, molecular, bioinformatics & 42\\
        \hline
        Engineering \& environment & simulation, energy, building, construction, engineering & 42\\
        \hline
        Clinical \& healthcare & clinical, healthcare, medical, patient, ai & 35 \\
        \hline
        Laboratory & scientific, laboratory, edor, microscopy, science & 22\\
        \hline
        Finance \& blockchain & financial, blockchain, market, trading, risk & 21\\
    \end{tabular}
\end{table}

In~\tab{tab:mcp-publication-contexts}, we present the distribution of MCP usage contexts. It indicates that most MCP-related publications do not treat MCP itself as their primary object of study. That is, relatively few publications focus on analyzing, extending, or evaluating the protocol. Instead, 57.8\% of publications use or apply MCP as an enabling technology to support a particular tool, workflow, or research objective.

\begin{table}[t]
    \centering
    \footnotesize
    \caption{Distribution of MCP usage contexts among academic publications and GitHub repositories.}
    \label{tab:mcp-publication-contexts}
    \begin{tabular}{l|rr}
        \textbf{usage context} & \textbf{publications (\%)} & \textbf{repositories (\%)}\\
        \hline
        \texttt{proposes\_extends}       & 7.9  & 1.5\\
        \texttt{studies\_analyzes}       & 7.9 & 1.2\\
        \texttt{evaluates\_benchmarks}   & 6.0 & 0.6\\
        \texttt{security\_privacy}       & 12.9 & 1.4\\
        \texttt{uses\_applies}           & 57.8 & 93.7\\
        \texttt{mentions\_only}          & 7.5 & 1.6 \\
    \end{tabular}
\end{table}

\begin{figure*}
    \centering
    \includegraphics[width=\linewidth]{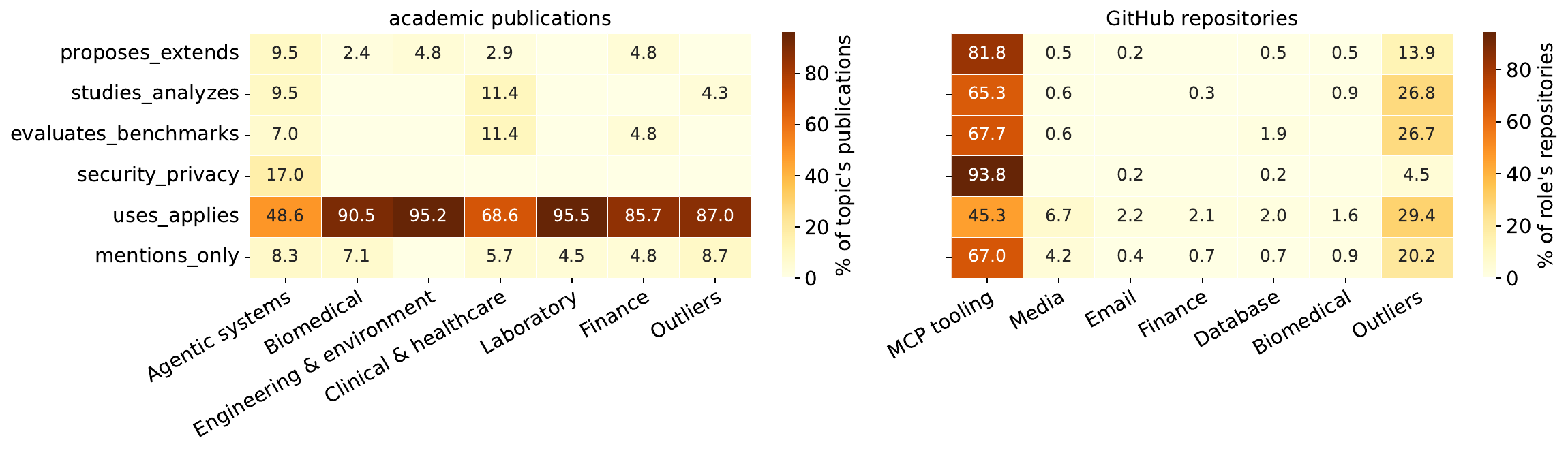}
    \caption{Distribution of \mcp usage contexts across application domains. Left: academic publications normalised by application domains. Right: GitHub repositories normalised by usage context.}
    \label{fig:heatmap}
\end{figure*}

\fig{fig:heatmap} (left) compares \mcp usage contexts across the identified application domains. Although publications classified as \texttt{uses\_applies} dominate across most domains, publications proposing or extending MCP are distributed relatively broadly. Further, publications that analyze MCP or develop benchmarks are particularly represented in the clinical and medical domain.

\begin{framed}
    \noindent\textbf{Finding 2.1:} Most MCP-related publications treat MCP as an enabling technology rather than the primary object of study. Studies analyzing or evaluating MCP are less common and occur mainly in clinical and healthcare research, with some evaluation work also appearing in finance and blockchain. Notably, security and privacy studies are absent from domain-specific topics.
\end{framed}

\subsubsection{\textbf{RQ2.2}: How are application domains and usage contexts distributed across \mcp-related GitHub repositories?}
\label{sec:rq22}

In our dataset of 33,319 repositories (\tab{tab:funnel}), we first restricted to repositories with English-language description/README text, retaining 31,058 repositories (93.2\%). 
We then restricted to repositories with at least 2,000 characters of combined description and README text, retaining 28,325 repositories (91.2\% of the English-language subset; 85.0\% of the full dataset).
The length threshold is consistent with Prana et al.'s approach \cite{prana2019categorizing}, who found that files below this length rarely contain substantive documentation.
Applying the BERTopic model on the description and README text of these 28,325 repositories yielded 15 topics ranging in size from \mcp tooling (13,393 repositories, 47.3\%) to Legal (246 repositories, 0.9\%) and one outlier bucket (8,111 repositories, 28.6\%).

Table~\ref{tab:repo_topics} reports the identified topic clusters, together covering 71.4\% of the dataset.
The remaining 28.6\% corresponds to the outlier bucket.
Among the 10 central terms provided by BERTopic, we report five that best represents the cluster.
\tab{tab:mcp-publication-contexts} (repositories column) reports the distribution of repositories among the usage context (\tab{tab:mcp-usage-taxonomy}), and we observe \texttt{uses\_applies} accounts for 93.7\% of the population.
This confirms that most repositories treat \mcp as an enabling technology rather than an object of study in its own right.
\begin{table}[t]
\centering
\footnotesize
\caption{Topics of \github repositories, five chosen description terms, and the number of repositories under each topic.}
\label{tab:repo_topics}
\begin{tabular}{p{2.1cm}|p{4.5cm}r}
\textbf{Topic} & \textbf{Central Terms} & \textbf{\# Repos} \\
\midrule
\mcp tooling         & ai, claude, memory, configuration, cli    & 13,393 \\
Media                & image, audio, video, unity, figma             & 1,791 \\
Email                & email, google, gmail, mail, imap              & 582 \\
Finance              & market, trade, stock, price, order            & 568 \\
Database             & database, sql, table, mysql, query            & 543 \\
Biomedical           & paper, clinical, gene, drug, fhir           & 423 \\
Browser              & browser, chrome, element, tab, dom      &  416 \\
Crypto payment       & wallet, x402, chain, mainnet, solana         &  394\\
Mobile app           & android, device, ios, xcode, adb       &  383\\
Web crawler          & search, content, web, crawl, scraping       &  331\\
Travel booking       & flights, booking, city, station, travel       &  330\\
EU AI Act Compliance & meok, eu ai, ai act, article, labs     &  310\\
Microsoft      & azure, microsoft, foundary, fabric, entra       &  257\\
Bi-lingual tooling   & tool, usage, configuration, ai, codex     &  247\\
Legal                & ansvar, legal, eu, regulatory, law       &  246
\end{tabular}
\end{table}

\fig{fig:heatmap} (Right) shows, for a given usage context, the share of its repositories in each of the top six major topics (Note: Rows do not sum to 100\%; the remainder corresponds to the nine minor topics (Table~\ref{tab:repo_topics}) omitted here for readability). 
From the heatmap, we observe that every usage context concentrates in \mcp tooling, ranging from 45.3\% for \texttt{uses\_applies}, the lowest, to 93.8\%, for \texttt{security\_privacy}.
Normalizing the same repositories by topic rather than usage context, a view not shown in \fig{fig:heatmap}, reveals that \texttt{uses\_applies} dominates every topic, from 89.9\% in \mcp tooling, the lowest, up to 100\% in Legal. 
These are near-monolithic ``build a server for X'' categories, wrapping a single existing application as an \mcp server (\eg \textit{mikhashev/law7} and \textit{durbs182/uk-tax-mcp}).
This shows repositories that evaluate, extend, secure, or study \mcp itself concentrate almost exclusively in the general-purpose \mcp tooling topic, which is the ecosystem's protocol-level core rather than a product category of its own. 
In contrast, repositories that merely apply \mcp are comparatively specialized in product-specific niches.

\begin{framed}
\noindent \textbf{Finding 2.2:} 93.7\% of \github repositories apply \mcp as an enabling technology, while 47.3\% fall under one dominant topic, \mcp tooling. 
Repositories that evaluate, extend, study, or secure \mcp concentrate almost exclusively in \mcp tooling (69.6 to 94.9\%), whereas \texttt{uses\_applies} repositories diversify into product-specific niches from Media to Legal (89.9 to 100\%).
\end{framed}

\subsubsection{\textbf{RQ2.3}: To what extent do application domains and usage contexts align between publications and repositories?}
\label{sec:rq23}

Both the publication and repository datasets fit independent topic models, yet surface overlapping domains, Biomedical and Finance, alongside one large general-purpose topic in each (Agentic systems for publications and \mcp tooling for repositories). 
Publications also cover the surface Engineering \& environment, Clinical \& healthcare and laboratory, while repositories cover the media, email and Database domains.
Usage contexts diverge more sharply, especially for \texttt{security\_privacy} (12.9\% vs. 1.4\%) as seen in \tab{tab:mcp-publication-contexts}.
Contexts other than \texttt{uses\_applies} together account for 42.2\% of publications versus 6.3\% of repositories, indicating non-\texttt{uses\_applies} is less common in repositories.
Each of the repository contexts individually concentrates within \mcp\ tooling at 45.3\% to 93.8\%, while each application domain individually concentrates within \texttt{uses\_applies}.

\begin{framed}
    \noindent\textbf{Finding2.3:} Publications and repositories converge on the same two domain-level topics, have different general purpose topics and diverges sharply in usage context.
\end{framed}
\subsection{Combining RQ1 and RQ2}
\label{sec:disc_rq1_rq2}

\subsubsection{Difference in observation between the two waves} GitHub's repository count and publication count differ by scale, as the dataset is 41 times larger (33,319 repositories versus 802 publications), a pattern documented for other fast-moving technical communities~\cite{saroar2024marketplace,salzano2025bridging}.
RQ1.2 showed that a second wave started in January 2026. 
Splitting each dataset in RQ2 as of 1 January 2026 resulted in 14,055 and 14,270 (50.7\%) repositories, and 381 and 392 (50.4\%) publications. 

The proportion of application domains is largely stable in aggregate for both datasets, though the underlying splits differ.
On \github, \mcp\ tooling's share of the repositories is essentially unchanged (47.2\% to 47.4\%).
Yet the five product-specific topics identified in RQ2.2 contract from 14.8\% to 12.8\% of the dataset, while the nine minor topics grow from 8.2\% to 12.3\%.  
Much of this growth concentrates in a single topic, EU AI Act compliance, which rises from 3 (0.0\%) to 307 (2.2\%) repositories across the wave boundary.
Publications take a different path as observed in RQ2.1.
Agentic systems fall from 80.3\% to 71.9\% of publications, and every established domain gains this share. For example, Biomedical from 3.7\% to 7.1\%, Engineering \& environment from 4.2\% to 6.6\%, and so on. 
Between these two waves, \github's new growth concentrated in one emerging niche while its established product categories contracted, whereas publications' new growth spread across existing domains.

Conversely, shifts in usage context proportion are not in the same direction. 
On the \github dataset, \texttt{security\_privacy}'s share grows 2.5$\times$ between waves (0.8\% to 2.0\%), while \texttt{proposes\_extends} falls to a quarter of its earlier share (2.4\% to 0.6\%). 
Publications show a different pattern. 
\texttt{security\_privacy}'s share is stable (13.1\% to 12.8\%), \texttt{evaluates\_benchmarks} nearly halves (7.9\% to 4.1\%), and \texttt{mentions\_only} grows by 1.6$\times$ (5.8\% to 9.2\%). 
This divergence is only visible once RQ1.2's wave boundary is applied to RQ2.3's usage-context comparison. 
As repositories and publications created close to the collection date have had less time to gain traction or be indexed, respectively, these figures should be read as provisional rather than settled.

\subsubsection{Common application domains} For the two application domains present in both datasets, Biomedical and Finance (\sect{sec:rq23}), repositories precede publications, though by very different margins. 
Excluding repositories created before \mcp's launch, using the same reasoning as RQ1.2 (\sect{sec:rq12}), Biomedical repositories reach a median creation date of 15 November 2025, over three months before Biomedical publications reach their median date of 25 February 2026, a 102-day gap. 
Finance shows a narrower gap of 23 days between repositories (9 November 2025) and publications (2 December 2025), close to the one-month lag RQ1.3 already reports between the two dimensions overall. 
With only two common domains present in both datasets, we read this as an observation specific to Biomedical and Finance rather than a general pattern of practice preceding research.

\subsubsection{Usage-contexts versus repository owner type and publication type}
\begin{table}[t]
\centering
\scriptsize
\caption{Usage context distribution (\%) by GitHub repository owner type and academic publication type, normalized within that owner type or publication type.}
\label{tab:role-by-ownertype-pubtype}
\begin{tabular}{@{}l|A{0.3cm}A{0.5cm}|A{0.95cm}A{0.4cm}A{0.4cm}A{0.7cm}@{}}
\textbf{Usage context} & \textbf{Org.} & \textbf{User} & \textbf{Book ch.} & \textbf{Conf.} & \textbf{Jour.} & \textbf{Prepr.} \\
\midrule
\texttt{proposes\_extends}     & 2.5  & 1.2  & 7.1  & 9.3  & 7.2  & 7.8  \\
\texttt{studies\_analyzes}     & 0.9  & 1.3  & 50.0 & 2.9  & 9.9  & 7.3  \\
\texttt{evaluates\_benchmarks} & 0.9  & 0.5  & 0.0  & 2.9  & 1.1  & 9.1  \\
\texttt{security\_privacy}     & 2.2  & 1.2  & 0.0  & 14.3 & 8.8  & 14.6 \\
\texttt{uses\_applies}         & 91.9 & 94.3 & 28.6 & 68.6 & 64.1 & 52.7 \\
\texttt{mentions\_only}        & 1.5  & 1.6  & 14.3 & 2.1  & 8.8  & 8.4  \\
\hline
\textbf{Absolute total}           & 6,777 & 21,548 & 14 & 140 & 181 & 438 \\
\end{tabular}
\end{table}

\texttt{Uses\_applies} share is nearly identical across \github owner types (Table~\ref{tab:role-by-ownertype-pubtype}), with organizational accounts showing a consistently higher, though modest, share of every protocol-level category. 
Publication type shows a sharper and more counter-intuitive pattern: conference proceedings and journal articles skew toward \texttt{uses\_applies} than preprints do. 
Preprints instead show the highest \texttt{evaluates\_benchmarks} share of the three well-populated types. 
Book chapters skew toward \texttt{studies\_analyzes} (50\%), consistent with the taxonomy's own inclusion of educational and survey material under this usage-context.
This pattern is based on a small group and should be read as illustrative rather than conclusive.

\section{Discussion}
\label{sec:discussion}

\subsection{Implications for practitioners}
\label{sec:dic_practitioners}
The five product-specific topics identified in RQ2.2 (\sect{sec:rq22}) are near-monolithic \texttt{uses\_applies} categories (97.9\% to 99.3\%), consistent with wrapping a single existing application as an \mcp server. 
This pattern echoes Mastouri et al.'s finding that 92\% of \mcp tools are implemented as bare API wrappers~\cite{mastouri2026api}. 
Practitioners adopting a third-party \mcp server for a specific product or service should therefore verify which operations it actually exposes rather than assume parity with the API or application it wraps.

Security- and privacy-focused work concentrates almost exclusively in the general-purpose \mcp\ tooling topic rather than in product-specific niches (RQ2.2), and \texttt{security\_privacy} accounts for only 1.4\% of repositories overall (Table~\ref{tab:mcp-publication-contexts}). 
Community security scrutiny has therefore reached sensitive-data domains such as Finance and Biomedical only marginally. 
Combined with Hasan et al.'s finding that existing vulnerability-detection patterns do not capture all \mcp-specific issues~\cite{hasan2026security}, practitioners deploying \mcp-based tools in regulated or sensitive-data domains should not assume the ecosystem has already produced domain-specific security guidance, and should apply their own review accordingly.

Publications that study or benchmark \mcp are also proportionally concentrated in the Clinical \& healthcare domain relative to other application domains identified in RQ2.1. 
Practitioners building \mcp-based tools for these use cases have published evaluation methodologies and benchmarks to draw on than practitioners building in domains without this concentration, and should consult the literature accordingly before designing their own evaluation from scratch.

\subsection{Implications for researchers}
\label{sec:dic_researchers}

MCP-related publications have grown sharply since early 2025 (\sect{sec:rq11}), suggesting the area is gaining visibility and may offer researchers a still-open field to establish significant contribution.
Still, this growth trend should be read cautiously, as the MCP domain is still young and such AI-related fields are known by their volatility.

\mcp's application reach extends well beyond computer science, spanning biomedical, engineering, clinical, and finance domains (\sect{sec:rq21}). 
Clinical \& healthcare, and to a lesser extent Finance \& blockchain, are the only domains where studying or benchmarking \mcp is proportionally well represented. 
In other domains, publications treat \mcp almost entirely as an enabling technology. 
This kind of study, evaluation, or security work is rare on \github, where it concentrates almost exclusively within the \mcp tooling topic (\sect{sec:rq22}) and is roughly seven times less common than in publications overall (6.3\% versus 42.2\%, \sect{sec:rq23}).

Future work should therefore move beyond demonstrating \mcp-based applications toward domain-specific evaluation and security research, particularly in sensitive-data domains such as healthcare and finance, where community scrutiny remains limited across both publications and repositories.
\section{Threats to validity}
\label{sec:threatstovalidity}

\subsubsection{Construct validity}

67 GitHub accounts each contribute at least 10 repositories, jointly accounting for 1,649 repositories (4.9\% of the corpus). The largest, \account{CSOAI-ORG}, contributes 295 repositories (0.9\%) alone. 
Manual inspection revealed these repositories could be generated from a template, and we found no principled basis for excluding them. 

Our repository-selection heuristic (\sect{sec:github_repo_data}) is grounded in prior work and an observed knee bend in corpus size. 
A stricter threshold would exclude more repositories without a proportional gain, and a looser one would retain more toy repositories. 
The manual validation (\sect{sec:github_repo_data}) found 96.8\% of repositories clearing this threshold to be genuine, bounding its practical impact.


Language identification relies on \texttt{langdetect}, which is generally accurate but not perfect, and the English-only filter does not guarantee monolingual content.
\textit{L-Qun/mcp-testing-framework} is one example of a repository with a bilingual README, and is retained as it also has an English version.

Our choice of topic size in BERTopic trades granularity for stability, which may merge sub-themes that emerge as distinct topics.
Clinical \& healthcare and Biomedical content form a single Biomedical topic in repositories but two separate topics in publications. 
We confirmed this merging is confined to a small number of low-frequency documents and does not change the dominant topics reported here. 

Only publications indexed by Lens.org containing the search term in their title, abstract, or keywords were included; publications relevant to \mcp\ outside this indexing or terminology may not have been captured.

\subsubsection{Internal validity}
Repositories created near the collection date have had less time to accumulate heuristics required to clear the thresholds. 
The decline in new repositories from March 2026 (\sect{sec:rq12}) is therefore a mix of genuine trend and this undercounting effect, which cannot be fully disentangled without a later observation. Similarly, for publications, conference papers may be indexed several months after the corresponding event, so publications presented between April and June 2026 may not yet have been indexed and are therefore absent from our corpus (\sect{sec:rq11}).

A rule-based data cleaning process was performed manually during the collection of research publications. This includes identifying duplicate records, correcting publication types and so on. 
Given the volume of documents analysed, residual errors in these fields cannot be ruled out.
\subsubsection{External validity}
This study characterizes \mcp's research and \github ecosystems as observed in June 2026. 
Given the pace of change already documented in RQ1 (\sect{sec:rq1}), the relative proportions reported here should be expected to shift at later snapshots.
The study also draws on exactly two data sources, \github and Lens.org, so findings may not generalize to other code-hosting platforms, such as GitLab, or other bibliographic databases, such as Scopus, which differ in coverage and community composition.
\subsubsection{Conclusion validity}
Our conclusions rest on quantitative, reproducible measurements throughout, and causal or generalizing language is avoided where there is lack of evidence. We are not aware of a specific threat to conclusion validity beyond those already discussed above.
\section{Conclusion and future work}
\label{sec:conclusion_future}

In this paper, we presented a longitudinal, two-dimensional study of \mcp's emergence in academic publications and adoption in \github repositories, analyzing 802 publications and 33,319 repositories that capture \mcp\ usage beyond established implementations. 
Despite a 41$\times$ difference in scale, both dimensions grew nearly synchronously, inflecting within about one month of each other in 2025 and peaking in March 2026. 
Growth had not flattened by the end of the observation window, with the 2026 wave accounting for 50.4\% of publications and 49.5\% of repositories.

Most publications (57.8\%) and repositories (93.7\%) treat \mcp as an enabling technology rather than an object of study. 
Publications that propose or extend \mcp span several domains, while studies and benchmarks are comparatively concentrated in Clinical \& healthcare. 
Repositories show a different pattern: repositories that study, evaluate, extend, or secure \mcp itself concentrate almost exclusively within a single general-purpose \mcp tooling topic, while repositories that apply \mcp diversify into product-specific niches such as email, finance, and biomedical tooling. 

Most importantly, security research has not followed \mcp's expansion into sensitive domains. Security- and privacy-focused publications appear only in the general-purpose Agentic systems topic, with none identified in Biomedical, Finance, Clinical \& healthcare, or other domain-specific topics. 
Similarly, security-focused repositories are rare across product-specific niches and absent from Media and Finance. 
Future work should therefore develop domain-specific security analyses, threat models, privacy assessments, and benchmarks for \mcp deployments involving sensitive data such as financial, clinical, and biomedical data.

In future, one can establish a direct link between publications and their corresponding repositories to test whether publication activity predicts a repository's subsequent traction. 
Another is to extend the same methodology to other domains to identify if the patterns observed in this paper are specific to \mcp or characteristic of fast-emerging technical ecosystems.

\bibliographystyle{IEEEtran}
\bibliography{biblio.bib}

\begin{thebibliography}{10}
\providecommand{\url}[1]{#1}
\csname url@samestyle\endcsname
\providecommand{\newblock}{\relax}
\providecommand{\bibinfo}[2]{#2}
\providecommand{\BIBentrySTDinterwordspacing}{\spaceskip=0pt\relax}
\providecommand{\BIBentryALTinterwordstretchfactor}{4}
\providecommand{\BIBentryALTinterwordspacing}{\spaceskip=\fontdimen2\font plus
\BIBentryALTinterwordstretchfactor\fontdimen3\font minus \fontdimen4\font\relax}
\providecommand{\BIBforeignlanguage}[2]{{%
\expandafter\ifx\csname l@#1\endcsname\relax
\typeout{** WARNING: IEEEtran.bst: No hyphenation pattern has been}%
\typeout{** loaded for the language `#1'. Using the pattern for}%
\typeout{** the default language instead.}%
\else
\language=\csname l@#1\endcsname
\fi
#2}}
\providecommand{\BIBdecl}{\relax}
\BIBdecl

\bibitem{antropic2024}
Antropic, ``Introducing the model context protocol,'' \url{https://www.anthropic.com/news/model-context-protocol}, 2024.

\bibitem{parra2025mcp}
D.~S. Parra, ``Mcp joins the agentic ai foundation,'' \url{https://blog.modelcontextprotocol.io/posts/2025-12-09-mcp-joins-agentic-ai-foundation/}, 2025.

\bibitem{hasan2026security}
M.~M. Hasan, H.~Li, E.~Fallahzadeh, G.~K. Rajbahadur, B.~Adams, and A.~E. Hassan, ``Model context protocol (mcp) at first glance: Studying the security and maintainability of mcp servers,'' \emph{Transactions on Software Engineering and Methodology}, 5 2026.

\bibitem{lin2025evolvable}
Z.~Lin, B.~Ruan, J.~Liu, and W.~Zhao, ``A large-scale evolvable dataset for model context protocol ecosystem and security analysis,'' in \emph{International Conference on Automated Software Engineering ({ASE})}, 2025, pp. 3985--3988.

\bibitem{mastouri2026api}
M.~Mastouri, E.~Ksontini, A.~Barrak, and W.~Kessentini, ``From rest to mcp: An empirical study of api wrapping and automated server generation for llm agents,'' 4 2026, preprint.

\bibitem{toeppe2026dataset}
B.~Toeppe, A.~Barrak, and E.~Ksontini, ``A large-scale dataset of mcp implementations on github,'' in \emph{International Conference on Mining Software Repositories ({MSR})}.\hskip 1em plus 0.5em minus 0.4em\relax Association for Computing Machinery ({ACM}), 4 2026.

\bibitem{saroar2024marketplace}
S.~K.~G. Saroar, W.~Ahmed, E.~Onagh, and M.~Nayebi, ``{GitHub} marketplace for automation and innovation in software production,'' \emph{Information and Software Technology}, vol. 175, p. 107522, 2024.

\bibitem{salzano2025bridging}
F.~Salzano, L.~Marchesi, C.~K. Antenucci, S.~Scalabrino, R.~Tonelli, R.~Oliveto, and R.~Pareschi, ``Bridging the gap: A comparative study of academic and developer approaches to smart contract vulnerabilities,'' 2025, preprint.

\bibitem{demetrescu2022computer}
C.~Demetrescu, I.~Finocchi, A.~Ribichini, and M.~Schaerf, ``On computer science research and its temporal evolution,'' \emph{Scientometrics}, vol. 127, no.~8, pp. 4913--4938, 2022.

\bibitem{dalle2022understanding}
M.~Dalle Lucca~Tosi and J.~C. dos Reis, ``Understanding the evolution of a scientific field by clustering and visualizing knowledge graphs,'' \emph{Journal of Information Science}, vol.~48, no.~1, pp. 71--89, 2022.

\bibitem{de2023global}
I.~M. De~la Vega~Hern{\'a}ndez, A.~S. Urdaneta, and E.~Carayannis, ``Global bibliometric mapping of the frontier of knowledge in the field of artificial intelligence for the period 1990--2019,'' \emph{Artificial Intelligence Review}, vol.~56, no.~2, pp. 1699--1729, 2023.

\bibitem{tosi2024metascience}
M.~D.~L. Tosi, J.~L.~C. Izquierdo, and J.~Cabot, ``A metascience study of the low-code scientific field,'' \emph{arXiv preprint arXiv:2408.05975}, 2024.

\bibitem{tosi2025ocl}
M.~D.~L. Tosi and J.~Cabot, ``Ocl on life support: can we revitalize the community for a stronger future?'' in \emph{STAF Workshops}, 2025.

\bibitem{oyelayo2026challenges}
O.~Oyelayo, S.~Abedu, S.~Khatoonabadi, and E.~Shihab, ``Developer challenges in the adoption of model context protocol in ai agent development.''\hskip 1em plus 0.5em minus 0.4em\relax Association for Computing Machinery ({ACM}), 4 2026, p. 53–59.

\bibitem{taraghi2026faults}
M.~Taraghi, M.~M. Morovati, and F.~Khomh, ``Real faults in model context protocol (mcp) software: a comprehensive taxonomy,'' 5 2026, preprint.

\bibitem{stein2026tools}
M.~Stein, ``How are ai agents used? evidence from 177,000 mcp tools,'' 3 2026, preprint.

\bibitem{escamilla2022rise}
E.~Escamilla, M.~Klein, T.~Cooper, V.~Rampin, M.~C. Weigle, and M.~L. Nelson, ``The rise of github in scholarly publications,'' in \emph{International Conference on Theory and Practice of Digital Libraries}.\hskip 1em plus 0.5em minus 0.4em\relax Springer, 2022, pp. 187--200.

\bibitem{wattanakriengkrai2022github}
S.~Wattanakriengkrai, B.~Chinthanet, H.~Hata, R.~G. Kula, C.~Treude, J.~Guo, and K.~Matsumoto, ``Github repositories with links to academic papers: Public access, traceability, and evolution,'' \emph{Journal of Systems and Software}, vol. 183, p. 111117, 2022.

\bibitem{alrashedy2024software}
K.~Alrashedy and A.~Binjahlan, ``How do software engineering researchers use github? an empirical study of artifacts \& impact,'' in \emph{2024 IEEE International Conference on Source Code Analysis and Manipulation (SCAM)}.\hskip 1em plus 0.5em minus 0.4em\relax IEEE, 2024, pp. 118--130.

\bibitem{fan2021makes}
Y.~Fan, X.~Xia, D.~Lo, A.~E. Hassan, and S.~Li, ``What makes a popular academic ai repository?'' \emph{Empirical Software Engineering}, vol.~26, no.~1, p.~2, 2021.

\bibitem{lens2026scholarly}
{The Lens}, ``Scholarly works search,'' \url{https://support.lens.org/knowledge-base/scholarly-works-search/}, 2026, documentation page accessed 8 June 2026.

\bibitem{munaiah2017curating}
N.~Munaiah, S.~Kroh, C.~Cabrey, and M.~Nagappan, ``Curating {GitHub} for engineered software projects,'' \emph{Empirical Software Engineering}, vol.~22, no.~6, pp. 3219--3253, 2017.

\bibitem{borges2018star}
B.~Hudson and T.~V. Marco, ``What’s in a github star? understanding repository starring practices in a social coding platform,'' \emph{Journal of Systems and Software}, vol. 146, pp. 112--129, 2018.

\bibitem{satopaa2011kneedle}
V.~Satopaa, J.~Albrecht, D.~Irwin, and B.~Raghavan, ``Finding a "kneedle" in a haystack: Detecting knee points in system behavior,'' in \emph{International Conference on Distributed Computing Systems Workshops}, 2011, pp. 166--171.

\bibitem{cochran1977sampling}
W.~G. Cochran, \emph{Sampling techniques}, 3rd~ed.\hskip 1em plus 0.5em minus 0.4em\relax John Wiley \& Sons, 1977.

\bibitem{jacob1960agreement}
J.~Cohen, ``A coefficient of agreement for nominal scales,'' \emph{Educational and Psychological Measurement}, vol.~20, no.~1, pp. 37--46, 1960.

\bibitem{gwet2008agreement}
K.~L. Gwet, ``Computing inter-rater reliability and its variance in the presence of high agreement,'' \emph{British Journal of Mathematical and Statistical Psychology}, vol.~61, no.~1, pp. 29--48, 2008.

\bibitem{feinstein1990high}
A.~R. Feinstein and D.~V. Cicchetti, ``High agreement but low kappa: I. the problems of two paradoxes,'' \emph{Journal of clinical epidemiology}, vol.~43, no.~6, pp. 543--549, 1990.

\bibitem{Gaddam_2024}
R.~R. Gaddam, ``Model context protocol ({MCP}) security and tenancy boundaries,'' \emph{International Journal of AI, BigData, Computational and Management Studies}, vol.~5, no.~1, p. 178–188, 3 2024.

\bibitem{mcp_apps_2026}
{Model Context Protocol}, ``Mcp apps,'' \url{https://blog.modelcontextprotocol.io/posts/2026-01-26-mcp-apps/}, 1 2026.

\bibitem{grootendorst2022bertopic}
M.~Grootendorst, ``{BERTopic}: Neural topic modeling with a class-based {TF-IDF} procedure,'' 2022, preprint.

\bibitem{wang2024identifying}
Z.~Wang, J.~Chen, J.~Chen, and H.~Chen, ``Identifying interdisciplinary topics and their evolution based on bertopic,'' \emph{Scientometrics}, vol. 129, no.~11, pp. 7359--7384, 2024.

\bibitem{liu2024unveiling}
Y.~Liu and F.~Wan, ``Unveiling temporal and spatial research trends in precision agriculture: A bertopic text mining approach,'' \emph{Heliyon}, vol.~10, no.~17, 2024.

\bibitem{zhao2024empirical}
Z.~Zhao, Y.~Chen, A.~A. Bangash, B.~Adams, and A.~E. Hassan, ``An empirical study of challenges in machine learning asset management,'' \emph{Empirical Software Engineering}, 2024.

\bibitem{alam2026analyzing}
K.~Alam and B.~Roy, ``Analyzing {GitHub} issues and pull requests in nf-core pipelines: Insights into nf-core pipeline repositories,'' in \emph{International Conference on Mining Software Repositories (MSR 2026)}, 2026.

\bibitem{izadi2021topic}
M.~Izadi, A.~Heydarnoori, and G.~Gousios, ``Topic recommendation for software repositories using multi-label classification algorithms,'' \emph{Empirical Software Engineering}, vol.~26, no.~5, p.~93, 2021.

\bibitem{nakatani2010langdetect}
S.~Nakatani, ``Language detection library for {Java},'' \url{http://code.google.com/p/language-detection/}, 2010.

\bibitem{reimers2019BERT}
N.~Reimers and I.~Gurevych, ``Sentence-{BERT}: Sentence embeddings using {S}iamese {BERT}-networks,'' in \emph{Empirical Methods in Natural Language Processing and the International Joint Conference on Natural Language Processing (EMNLP-IJCNLP)}.\hskip 1em plus 0.5em minus 0.4em\relax Association for Computational Linguistics, 11 2019, pp. 3982--3992.

\bibitem{song2020mpnet}
\BIBentryALTinterwordspacing
K.~Song, X.~Tan, T.~Qin, J.~Lu, and T.-Y. Liu, ``Mpnet: masked and permuted pre-training for language understanding,'' in \emph{International Conference on Neural Information Processing Systems}.\hskip 1em plus 0.5em minus 0.4em\relax Curran Associates Inc., 2020. [Online]. Available: \url{https://dl.acm.org/doi/10.5555/3495724.3497138}
\BIBentrySTDinterwordspacing

\bibitem{mcinnes2018umap}
L.~McInnes, J.~Healy, N.~Saul, and L.~Gro{\ss}berger, ``Umap: Uniform manifold approximation and projection,'' \emph{Journal of Open Source Software}, vol.~3, no.~29, p. 861, 2018.

\bibitem{campello2013hdbscan}
R.~J. G.~B. Campello, D.~Moulavi, and J.~Sander, ``Density-based clustering based on hierarchical density estimates,'' in \emph{Advances in Knowledge Discovery and Data Mining}, ser. Lecture Notes in Computer Science, vol. 7819.\hskip 1em plus 0.5em minus 0.4em\relax Springer, 2013, pp. 160--172.

\bibitem{gilardi2023chatgpt}
F.~Gilardi, M.~Alizadeh, and M.~Kubli, ``Chatgpt outperforms crowd workers for text-annotation tasks,'' \emph{Proceedings of the National Academy of Sciences}, vol. 120, no.~30, p. e2305016120, 2023.

\bibitem{prana2019categorizing}
G.~A.~A. Prana, C.~Treude, F.~Thung, T.~Atapattu, and D.~Lo, ``Categorizing the content of {GitHub} {README} files,'' \emph{Empirical Software Engineering}, vol.~24, no.~3, pp. 1296--1327, 2019.

\end{thebibliography}

\end{document}